\documentclass[a4paper,12pt]{article}
\usepackage{amsfonts,amsthm,amsmath,amssymb,graphicx,color,hyperref,youngtab}
\usepackage[bulletsep]{collref}
\usepackage[lmargin=3.0cm, rmargin=1.5cm,tmargin=2.50cm,bmargin=2.50cm]{
geometry}
\usepackage{mathrsfs}
\usepackage{float}
\usepackage{ulem}

\newcommand{\be}{\begin{equation}}
\newcommand{\ee}{\end{equation}}
\newcommand{\bea}{\begin{eqnarray}}
\newcommand{\eea}{\end{eqnarray}}

\newcommand{\nn}{\nonumber}
\newcommand{\Tr}{\text{Tr}}
\newcommand{\mL}{\mathcal{L}}

\begin{document}


\def\gap#1{\vspace{#1 ex}}
\def\be{\begin{equation}}
\def\ee{\end{equation}}
\def\bal{\begin{array}{l}}
\def\ba#1{\begin{array}{#1}}  
\def\ea{\end{array}}
\def\bea{\begin{eqnarray}}
\def\eea{\end{eqnarray}}
\def\beas{\begin{eqnarray*}}
\def\eeas{\end{eqnarray*}}
\def\del{\partial}
\def\eq#1{(\ref{#1})}
\def\fig#1{Fig \ref{#1}} 
\def\re#1{{\bf #1}}
\def\bull{$\bullet$}
\def\nn{\\\nonumber}
\def\ub{\underbar}
\def\nl{\hfill\break}
\def\ni{\noindent}
\def\bibi{\bibitem}
\def\ket{\rangle}
\def\bra{\langle}
\def\vev#1{\langle #1 \rangle} 
\def\lsim{\stackrel{<}{\sim}}
\def\gsim{\stackrel{>}{\sim}}
\def\mattwo#1#2#3#4{\left(
\begin{array}{cc}#1&#2\\#3&#4\end{array}\right)} 
\def\tgen#1{T^{#1}}
\def\half{\frac12}
\def\floor#1{{\lfloor #1 \rfloor}}
\def\ceil#1{{\lceil #1 \rceil}}

\def\mysec#1{\gap1\ni{\bf #1}\gap1}
\def\mycap#1{\begin{quote}{\footnotesize #1}\end{quote}}

\def\Om{\Omega}
\def\b{\beta}
\def\s{\sigma}

\def\lan{\langle}
\def\ran{\rangle}

\def\bit{\begin{item}}
\def\eit{\end{item}}
\def\benu{\begin{enumerate}}
\def\eenu{\end{enumerate}}

\makeatletter
\@addtoreset{equation}{section}
\makeatother
\renewcommand{\theequation}{\thesection.\arabic{equation}}

\rightline{TIFR/TH/13-16}
\rightline{WITS-CTP-116}
\vspace{1.2truecm}

\vspace{1pt}


{\Large{
\begin{center}{\bf Dynamical entanglement entropy 
with angular momentum and U(1) charge}
\end{center}
}}

\vskip.9cm

\thispagestyle{empty} \centerline{\large \bf  
Pawe{\l} Caputa$^{a,b}$\,\footnote{pawel.caputa@wits.ac.za}, Gautam Mandal$^{b}$\,\footnote{mandal@theory.tifr.res.in} and Ritam Sinha$^{b}$\,\footnote{ritam@theory.tifr.res.in}}

\vspace{.8cm}
\centerline{$^a${\it NITheP, Department of Physics and Centre for Theoretical Physics}}
\centerline{{\it University of the Witwatersrand, Wits, 2050, South Africa } }
\vspace{.2cm}
\centerline{$^b${\it  Department of Theoretical Physics}}
\centerline{{\it Tata Institute of Fundamental Research, Mumbai 400005, India.} }


\gap7


\gap3

\thispagestyle{empty}

\gap6

\centerline{\bf Abstract}
\vskip.5cm 

We consider time-dependent entanglement entropy (EE) for a 1+1
dimensional CFT in the presence of angular momentum and U(1) charge.
The EE saturates, irrespective of the initial state, to the grand
canonical entropy after a time large compared with the length of the
entangling interval. We reproduce the CFT results from an AdS dual
consisting of a spinning BTZ black hole and a flat U(1) connection. The
apparent discrepancy that the holographic EE does not {\it a priori}
depend on the U(1) charge while the CFT EE does, is resolved by the
charge-dependent shift between the bulk and boundary stress
tensors. We show that for small entangling intervals, the entanglement
entropy obeys the first law of thermodynamics, as conjectured
recently. The saturation of the EE in the field theory is shown to
follow from a version of quantum ergodicity; the derivation indicates
that it should hold for conformal as well as massive theories in any
number of dimensions.

\setcounter{page}{0}
\setcounter{tocdepth}{2}

\newpage

\tableofcontents

\section{\label{sec:intro}Introduction and Summary}

Entanglement entropy (EE) of a quantum system has turned out be a
useful observable in many areas of physics; see reviews \cite{EE,Nishioka:2009un,Calabrese:2009qy,Eisert:2008ur}. In this paper, we
will primarily use EE as a dynamical tool, especially to describe
equilibration in 1+1 dimensional quantum field
theories. Time-dependent EE in 1+1 dimensional CFT has been studied in
detail in  \cite{Calabrese:2005in,Calabrese:2009qy}. Let us consider a CFT
with an infinite spatial direction; the EE for a single interval of
length $l$, is found to saturate, according to the formula 
\footnote{\label{ftnt:beta}In
  \cite{Calabrese:2005in,Calabrese:2009qy}, the entropy density is
  given by $s_{\rm eqm}= c\pi/(3\b)$, where $\beta$ is the inverse
  temperature of a canonical ensemble equivalent to a microcanonical
  ensemble at energy E, given by $\b= \sqrt{\pi c/6 E}$. With this, we
  recover the RHS of \eq{thermal}.}
\be 
S_{\rm  ent}(t,l| \psi ) \xrightarrow{t \gg l} l\, s_{\rm eqm}(E), \quad
s_{\rm eqm}(E)= \sqrt{2\pi c E/3}
\label{thermal}
\ee
In the above equation, the LHS is the EE of an interval of length $l$,
computed in the state $| \psi \rangle$ at time $t$; we will denote the
energy density of the state as $E$. $s_{\rm eqm}(E)$ is the
equilibrium entropy density in the microcanonical ensemble as a
function of energy density $E$. It is assumed here that length of the
interval $l$ is greater than the characteristic length scale $1/\sqrt
E$ associated with the state $| \psi \ran$ (this condition will play
an important role in Section \ref{sec:massive}).
\cite{Hartman:2013qma} showed that the time-development in
\eq{thermal} can be interpreted holographically in terms of a BTZ
black hole, and derived \eq{thermal} using the Ryu-Takayanagi
definition of holographic EE \cite{Ryu:2006bv} (see 
\cite{Balasubramanian:2011at,AbajoArrastia:2010yt,Albash:2010mv,Liu:2013iza}
for other recent
works on holographic thermalization using dynamic EE). The linear
growth in time was given an intuitive explanation in  terms
of oppositely moving entangled pair of excitations  \cite{Calabrese:2005in};
one of the objectives of this paper is to explain the saturation value 
in terms of quantum ergodicity. 

\medskip
An important point to note about \eq{thermal} is the {\it information
  loss} aspect of this equation: on the RHS of \eq{thermal}, all
information about the specific state $| \psi \ran$ appears to be lost,
other than energy $E$ of the state. We will elaborate on this further
in Section \ref{sec:info-loss}. Indeed, 
the  above statement of equilibration is similar in spirit to the
following statement of quantum ergodicity (see 
\cite{Polkovnikov:2010yn}, Section III-A)
\be
\Tr \left(\rho_{\rm pure} O \right)   \xrightarrow{t\to \infty}
\Tr \left( \rho_{\rm mc} O \right),  \quad   \rho_{\rm pure}
= | \psi \ran \lan \psi |, \quad  \rho_{\rm mc} =
\frac{1}{\Om(E)} \sum_{i \in {\cal H}_E} | i \ran \lan i |,
\label{ergodic}
\ee
which is believed to be true for a class of ``macroscopic''
observables $O$.  Here, $\rho_{\rm mc}$ defines a microcanonical
ensemble at energy $E$; ${\cal H}_E$ denotes the subspace
of states with this energy, and $\Om(E)$ is the dimension of
${\cal H}_E$.  We will, in fact, derive \eq{thermal} from
\eq{ergodic}, modulo some assumptions, in Section \ref{sec:erg}.  Thus,
in the time-development described in \eq{thermal} not only is the
memory of the initial state lost, the RHS is given in terms of a mixed
state. It is worth noting that in \eq{ergodic} no mention is made
about the time scale of change; thus, the time-development in 
\eq{thermal} provides a time scale for equilibration.

\medskip

In this paper, we will show that the above statement of equilibration
also holds in the presence of additional conserved charges, besides
the energy $E$. In particular, we will show that if the initial state
$| \psi \ran$ has a non-zero angular momentum $J$ 
\footnote{\label{ftnt:J}In this
paper, we will mostly be concerned with a non-compact spatial 
direction, so $J$ is actually a linear momentum. However, we
regard this non-compact direction as arising in the limit of a large circle 
(the bulk dual is a BTZ black string which can be regarded as the
limit of a BTZ black hole), and will continue to call $J$ an
`angular momentum'.} and a U(1) charge
$Q$, we have \footnote{The divergent piece $S_{\rm div}$ is the same
as in the previous literature, including in 
\cite{Calabrese:2005in,Hartman:2013qma}. We do not have
anything new to add regarding this term; for a recent
discussion, see  \cite{Liu:2012eea}.}
\begin{align}
&S_{\rm ent}(t,l| \psi ) \xrightarrow{t \gg l} l\, 
s_{\rm eqm}(E, J, Q)
\label{thermal-chem}
\end{align}
Here $s_{\rm eqm}(E, J, Q)$ equals the equilibrium entropy density in
a microcanonical ensemble, described in Eqs.\eq{s-micro-J} and
\eq{s-micro-JQ}, which give the detailed form of the time-dependence. 
Eq. \eq{thermal-chem} is the main result of our paper; it
is presented here in the limit in which the excitation energy of the
initial state is much higher than $1/t, 1/l$. The precise version of
this statement as well as the exact expression for the LHS without
this restriction is given in Sections \ref{CFT-J} and
\ref{CFT-JQ}). We derive \eq{thermal-chem} also from a holographic
set-up (Sections \ref{sec:hol-J} and \ref{sec:AdS-JQ}).  The
holographic dual consists of a spinning BTZ black hole plus a U(1)
gauge field described by a CS theory.

Eq. \eq{thermal-chem} leads us to the following natural conjecture for
an integrable 1+1 dimensional CFT.  Suppose the initial state
$|\psi\ran$ has an infinite number of non-zero conserved charges $Q_i,
i=1,..., \infty$ (including energy). We conjecture that the long time 
behaviour of the EE in this case is given by 
\begin{align} 
&S_{\rm ent}(t,L| \psi ) \xrightarrow{t \gg l} l\, s_{\rm eqm}(\{Q_i\})  
\nonumber\\
&s_{\rm eqm}(\{Q_i\})= s_{\rm GGE}(\{\mu_i\})
\label{thermal-GGE}
\end{align}
where $\mu_i$ are values of chemical potentials conjugate to the
infinite number of charges $Q_i$ carried by the quantum state $| \psi
\ran$. In the second line, we have used the equivalence between the
microcanonical ensemble (with infinite number of charges) and the
generalized Gibbs ensemble (GGE).  The corresponding generalization of
\eq{ergodic} to 2-dimensional integrable systems has already been
proved \cite{Rigol:2007, calabrese:2012, Mandal:2013id}.  A natural
speculation about a holographic dual of the above involves higher spin
black holes \cite{Ammon:2012wc} (see Section \ref{sec:hs} for a brief
discussion).

\medskip
A few remarks are in order:

\begin{enumerate}

\item We reproduce the CFT results in this paper from AdS in two ways:
  (a) by an explicit evaluation of the Ryu-Takayanagi (RT) formula for
  holographic EE and matching with CFT, and (b) showing that the RT
  formula follows from a conventional AdS/CFT dual of the CFT
  correlators of twist fields in a double scaling limit. Method (b)
  constitutes a `proof' of the RT prescription for 1+1 dimensional CFT
  (see Section \ref{sec:proof-ryu} for details).

\item We encounter a puzzle in applying the RT prescription for
  holographic EE in the presence of the U(1) charge. The U(1) charge
  we consider is implemented in the AdS dual by a U(1) Chern-Simons
  (CS) theory, and addition of a U(1) charge does not change the
  metric. Therefore the RT holographic EE is independent of the U(1)
  charge, which seems to be in conflict with the CFT expressions which
  clearly depend on this charge. We resolve this puzzle in Section
  \ref{sec:puzzle}.

\end{enumerate}

This work was first presented in the Seventh Crete regional meeting in
string theory \cite{GM-Crete:2013}. After the work was complete and it
was ready to be submitted, the preprints
\cite{deBoer:2013vca,Ammon:2013hba} appeared which have some
overlap with our paper, especially in Sections  \ref{sec:AdS-JQ}
and \ref{sec:Limits}.

\section{\label{CFT-J}Entanglement Entropies with spin: CFT}

\subsection{\label{CFT-J-double}Thermofield double}

Let us consider a 2D CFT at a finite temperature, which is represented
by a thermofield double consisting of two identical copies of the CFT
\cite{Takahashi:1975}. Consider the following initial (pure)
state, belonging to the thermofield double, on the time slice $t=0$:
\be
| \psi \ran = C \sum_i  \exp[- \beta  (E_i + \Om J_i)/2]\, 
| i\ran \otimes  | i \ran
\label{hartle-j}
\ee 
The normalization constant $C$ is given in terms of the partition function
\be
|C|^{-2}   \equiv Z(\beta, \Om)  \equiv 
\Tr\, e^{-\beta(H+\Om J)}= \Tr  \exp[-\beta_+ L_0 - \beta_- \bar L_0],
\label{partition}
\ee
and we use the following identifications
\be
H = L_0+\bar L_0,\qquad J=L_0-\bar{L}_0,\qquad
\b_\pm = \b (1 \pm \Om)
\label{b-pm}
\ee
The index $i$ of the sums goes over a complete set of states of $H_1$
(equivalently $H_2$). The definition implies that the expectation
values $\lan E \ran, \lan J \ran$ in this state are non-zero, and are
related to the inverse temperature $\b$ and the `angular velocity'
$\Om$ (which is essentially a chemical potential for the conserved
angular momentum).\footnote{As mentioned in footnote \ref{ftnt:J}, for
  a non-compact spatial direction, $J$ is actually a linear momentum;
$\Om$ is the corresponding chemical potential.} In a Euclidean spacetime,
$\Om$ must be chosen to be purely imaginary:  
\be
\Om = i \Om_E
\label{om-e}
\ee 
Now, consider two identical entangling regions $A\subset
\mathbb{R}$ in both copies of the field theory and compute the time
evolution of the EE. First, we take $A$ to be a semi-infinite
line. Following the prescription in \cite{Calabrese:2004eu} (see
review \cite{Calabrese:2009qy}) the entanglement Renyi entropy (ERE)
is given by the CFT functional integral over a Riemann surface
obtained by gluing $n$ copies of a cylinder \footnote{\label{ftnt:z}We
  will use complex coordinates $(z, \bar z)= \s_1 \pm i \s_2$ on the
  cylinder, with $-H$ and $iP$ as the generator of translation along
  $\s_2$ and $\s_1$. The formula \eq{partition} implies the following
  twisted identification $(\s_1, \s_2) \equiv (\s_1 - \b \Om_E, \s_2 +
  \b)$; in terms of the complex coordinates $z \equiv z + i\b_+$,$
  \bar z \equiv \bar z- i \b_-$, where we have used \eq{b-pm} and
  \eq{om-e}.}  along two semi-infinite cuts from $z_1=\bar{z}_1=0$ and
from $z_2=i\beta_+/2$, $\bar{z}_2=-i\beta_-/2$, both running off to
infinity (see Figure 1).  Such a partition function, in turn, 
boils down to the two-point function of twist fields $\phi_\pm$ 
\cite{Calabrese:2004eu} at the two branch points. Thus, the ERE is given by
\be
S^{(n)}= \frac{1}{1-n}\log \langle \phi^{+}(z_1,\bar{z}_1)
\phi^{-}(z_2,\bar{z}_2)\rangle 
\label{ere}
\ee

\begin{figure}[H]
\centerline{\includegraphics[width=400pt, height=120pt]{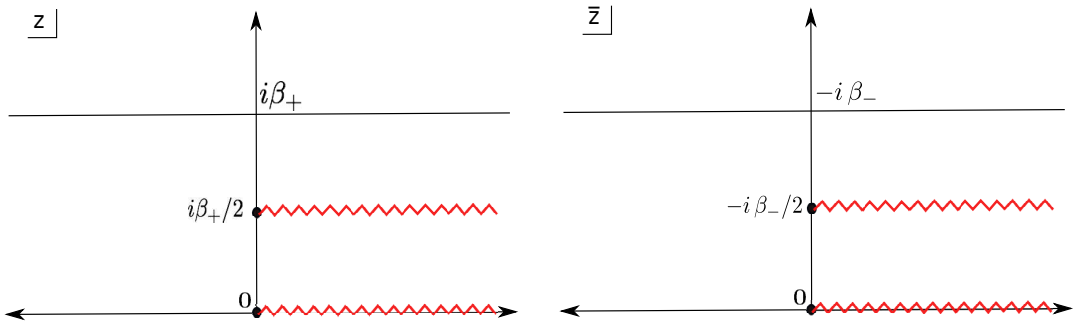}}
\caption{The lower branch cuts on left and right represent
the  (holomorphic and antiholomorphic coordinates) of 
the entangling interval in the first copy of the CFT. The upper branch
cuts represent the second CFT.}
\label{cylinder-fig}
\end{figure}

To obtain this two-point function we first map the
cylinder to the plane with coordinates given by
\be
w(z)=\exp\left(\frac{2\pi\, z}{\beta_+} \right),\qquad 
\bar{w}(\bar{z})=\exp\left(\frac{2\pi\, \bar{z}}{\beta_-} 
\right)\label{MapJ}
\ee
The two-point function on the plane of an operator $O$ 
with conformal dimensions
$(h, \bar h)$ is given by
\be
\langle O(w_1,\bar{w}_1)O(w_2,\bar{w}_2)\rangle=
\frac{1}{(w_2-w_1)^{2h}(\bar{w}_2-\bar{w}_1)^{2\bar{h}}}
\label{plane}
\ee
Now, under a conformal mapping $(w,\bar w) \to (z, \bar z)$,
correlators transform as 
\be
\langle O(z_1,\bar{z}_1)O(z_2,\bar{z}_2)...\rangle=\prod_i 
\left(\frac{dw_i}{dz_i}\right)^h
\left(\frac{d\bar w_i}{d\bar z_i}\right)^{\bar h} \langle 
O(w_1,\bar{w}_1)O(w_2,\bar{w}_2)...\rangle
\label{corr-trans}
\ee
The ERE \eq{ere} can be obtained by
using these results and the fact that for the twist fields $\phi_\pm$ of
order $n$
\be
h=\bar{h}=\frac{c}{24}\left(n-\frac{1}{n}\right)
\ee
As explained in \cite{Calabrese:2004eu}, the EE is obtained
by taking the $n \to 1$ limit. This gives  
\be
S_{EE}= S^{(1)}= \frac{c}{6} \log 
\left(\frac{\beta_+\beta_-}{\pi^2\epsilon^2}  
\sinh\frac{\pi (z_2 - z_1)}{\beta_+}
\sinh\frac{\pi (\bar z_2 - \bar z_1)}{\beta_-}\right)
\label{ee-val}
\ee
The cut-off $\epsilon$ is used to regularize the expression, as
in \cite{Calabrese:2004eu}.  \footnote{This equation appears 
in Ref. \cite{Hubeny:2007xt}, where it signifies the
equilibrium EE of a finite interval of length $|z_2-z_1|$.} 
To explicitly evaluate \eq{ee-val}
we substitute the values of $(z_{1,2}, \bar z_{1,2})$ mentioned
above \eq{ere}, and obtain the divergent value
\be
S_{EE}= S_{EE,0}= \frac{c}{6} \log 
\left(\frac{\beta_+\beta_-}{\pi^2\epsilon^2}  \right)
\label{ee-t=0}
\ee 
Here the subscript zero indicates that the EE is computed at
$t=0$.

\subsubsection{Time-dependent EE}

We will now  consider the (Lorentzian) time-evolution of the 
thermofield state \eq{hartle-j}:
\be
| \psi(t) \ran = \exp[-iHt]|\psi\ran 
= C \sum_i  \exp[- \beta  (E_i + \Om J_i)/2 -  i 2 E_i t]\, 
| i\ran \otimes  | i \ran
\label{hartle-j-t}
\ee 
and will compute the time-dependent ERE and EE based on 
this time-dependent state. In the notation of footnote \ref{ftnt:z}, 
the total evolution
operator in \eq{hartle-j-t} translates $(\s_1, \s_2)=(0,0)$ $\to$ $(\s_1, \s_2)
= (-\b\Om_E/2, \b/2 + 2 i t)$. This implies the following
analytically continued location of the two branch points     
\be
z_1=\bar{z}_1=0,\quad z_2=-2t+i\,\beta_+/2,\quad \bar{z}_2=2t-i\,\beta_-/2
\label{2-branch-pts}
\ee
Note that $\bar z_2 \ne z_2^*$; this happens because $\s_2$ is now complex.
By using the new locations of the
branch points in \eq{ee-val}
\be
S_{EE}= \frac{c}{6} \log
\left(\frac{\beta_+\beta_-}{\pi^2\epsilon^2}
\cosh\frac{2\pi t}{\beta_+}
\cosh\frac{2\pi t}{\beta_-}\right)
\ee
Clearly at large $t \gg \b, \b\Om_E$, the $\cosh$ terms
can be replaced by exponentials, which show that the finite part
grows linearly with time:
\begin{align}
& S_{EE}(t)= S_{EE,0} +  t\, (2 s_{\rm eqm}),
\quad s_{\rm eqm} = \frac{\pi c}{3\beta(1-\Omega^2)}  
\qquad t \gg  \b, \b\Om_E 
\label{s-micro-J-infinite}
\end{align}
where $s_{\rm eqm}$ is the equilibrium entropy density,
further elaborated below \eq{s-micro-J}. $S_{EE,0}$ 
is already defined in \eq{ee-t=0}.

\subsubsection{Finite interval}

Now, take $A$ be a finite interval of length $l$. In this case, we
need to consider a functional integral over the cylinder with two {\it
  finite} cuts. The locations of the branch points $(z_i, \bar z_i)$
are
\begin{align} 
& z_1=\bar{z}_1=0, \; z_2=\bar{z}_2=l, 
\nonumber\\ 
& z_3=l-2t+i\frac{\beta_{+}}{2},\;
\bar{z}_3=l+2t-i\frac{\beta_{-}}2,\quad 
z_4=-2t+i\frac{\beta_{+}}2,\; \bar{z}_4=2t-i\frac{\beta_{-}}2
\label{z-config}
\end{align}
As described in \cite{Calabrese:2005in,Hartman:2013qma}, the
entanglement Renyi entropy is given by the four-point correlator of
the twist fields
\be
S_{n}=\frac1{1-n}\log
\langle \phi^{+}(z_1,\bar{z}_1) \phi^{-}(z_2,\bar{z}_2) \phi^{+}(z_3,\bar{z}_3)
\phi^{-}(z_4,\bar{z}_4) \rangle
\label{CorrIn}
\ee
As before, a way to compute this would be by mapping the
points  to the plane using \eqref{MapJ}, computing the correlator
there and transforming back to the cylinder by using \eq{corr-trans}.
The details of this calculation are similar to the $\Om=0$ case discussed
in \cite{Hartman:2013qma}. 
The 4-point function on the plane depends on the cross-ratio
\be x=\frac{w_{12}w_{34}}{w_{13}w_{24}}=\frac{2\sinh^{2}\frac{\pi
    l}{\beta_+}}{\cosh\frac{2\pi
    l}{\beta_+}+\cosh\frac{4\pi\,t}{\beta_+}},\qquad
\bar{x}=\frac{\bar{w}_{12}\bar{w}_{34}}{\bar{w}_{13}\bar{w}_{24}}=
\frac{2\sinh^{2}\frac{\pi
    l}{\beta_-}}{\cosh\frac{2\pi
    l}{\beta_-}+\cosh\frac{4\pi\,t}{\beta_-}} \ee
where $w_{ij}=w(z_i)-w(z_j)$ and similarly for $\bar{w}$. Let us
assume that $l, t \gg \b, \b \Om_E$. We then have: $x \sim (1 +
\exp[\frac{4\pi}{\b_+}(t- l/2)])^{-1} $ up to $O(\exp[-t/\b_+],
\exp[-l/\b_+])$.  \\

Case (i): For $t < l/2$ \footnote{\label{ftnt:rounding}Strictly
  speaking, we need here $(l/2 - t) \gg \b_\pm$, to ensure $x\to 1$.},
we then have $x \to 1$, which, in terms of the original coordinates,
implies $z_2 \to z_3$ and hence a factorization $\lan 1~4 \ran \lan
2~3 \ran$. Once we realize this, we can go back to \eq{CorrIn} and
evaluate the four-point function as
\[
\langle \phi^{+}(z_1,\bar{z}_1) \phi^{-}(z_4,\bar{z}_4) \rangle
\langle \phi^{-}(z_2,\bar{z}_2) \phi^{+}(z_3,\bar{z}_3)\rangle
\]

Case (ii): For $t> l/2$ \footnote{We actually need $(t-l/2) \gg \b_\pm$.
See footnote \ref{ftnt:rounding}.}, by similar reasonings, we have
$x\to 0$, which implies the other factorization for the 4-point
function
\[
\langle \phi^{+}(z_1,\bar{z}_1) \phi^{-}(z_2,\bar{z}_2) \rangle
\langle \phi^{+}(z_3,\bar{z}_3)   \phi^{-}(z_4,\bar{z}_4) \rangle
\]
Using our results from the previous subsections about the two-point 
function, we find the following behaviour of the EE:
\begin{align}
&S_{EE}=\left \{       
\begin{array}{l l} 2t\, (2 s_{\rm eqm})
   +S_{\rm div} & \quad t\le l/2\\
\\ l\, (2 s_{\rm eqm})  
    +S_{\rm div}  & \quad t \ge l/2
  \end{array}          
                   \right.
\nonumber \\ 
& s_{\rm eqm} = \frac{\pi c}{3\beta(1-\Omega^2)}
=  \sqrt{\frac{\pi c}{6}(E+J)} + 
\sqrt{\frac{\pi c}{6}(E-J)}
\label{s-micro-J}
\end{align}
Clearly the EE saturates after time $t=l/2$. Here $s_{\rm eqm}$ is the
equilibrium entropy density of (either copy of) the CFT; the first
expression on the second line gives its value in the canonical
ensemble and the second expression gives the microcanonical value (see
Section \ref{sec:thermo} for the relation between the two
ensembles). For $\Om=0$ the results of this section correctly reduce
to those derived in
\cite{Hartman:2013qma}. \footnote{\label{ftnt:finite-double} The
  expression for the time-dependent entanglement entropy for the
  finite interval, in \eq{s-micro-J}, is twice that for the half-line
  \eq{s-micro-J-infinite}, as observed in \cite{Hartman:2013qma} for
  $\Om=0$. We will encounter the same feature for a single CFT, as
  well as for the bulk duals.} The divergent part $S_{div}$ is the
same as $S_{EE,0}$ is \eq{ee-t=0} and \eq{s-micro-J-infinite}.

We have thus proved \eq{thermal-chem} starting from a rather special
pure state of the form \eq{hartle-j}.  We will now present a more
general derivation starting from an {\it arbitrary} initial state.

\subsection{\label{sec:CFT-J-single}Single CFT, arbitrary state}

Let us now consider a single CFT, defined on a cylinder (with
coordinates described in footnote \ref{ftnt:z}). We start with a
pure state $| B \ran$ at time $\s_2=0$, and evolve it by (i)
translating in $\s_2$ by $\b/4$ (as in \cite{Hartman:2013qma}) 
and (ii) in $\s_1$ by $-\b \Om_E$; this leads to another pure state
\be
|\psi \ran = \exp[-\beta(H + \Om J)/4]| B \ran,
\label{b-state-eq}
\ee We will regard this as the initial state for further, Lorentzian,
time evolution, and compute the time-dependent EE for a single
interval in the state
\be
|\psi(t) \ran = \exp[-i H t] |
\psi \ran =  \exp[-(\b/4 + it)H - \b\Om J/4] | B \ran. 
\label{complex-t-single}
\ee
By choosing $| B \ran$ arbitrarily, we can obtain an
arbitrary initial state $| \psi \ran$. We will comment in Section
\ref{sec:info-loss} on the independence of the EE with respect to the
choice of this initial state.

Let us first consider the case where the interval is a half-line.
Suppose at $\s_2=0$, the half-line ends at $\s_1=0$. Then
after the evolution, this point is translated to 
$\s_1=-\b\Om_E/4, \s_2= \b/4+it$ \footnote{Recall that in \eq{complex-t-single},
$-H, iJ$ are, respectively, the translation operators in $\s_2, \s_1$,
and $\Om=i \Om_E$. See below \eq{hartle-j-t}.}, or, in terms of the 
$z, \bar z$ coordinates (see footnote \ref{ftnt:z}), to the point
\be
z_1=t+i\beta_+/4,\qquad \bar{z}_1=-(t+i\beta_-/4)
\label{z1-single}
\ee
The computation of the time-dependent EE, in part similar to that 
described above, involves a generalization of the techniques
in \cite{Calabrese:2005in, Hartman:2013qma}. The entanglement Renyi
entropy involves computing the one-point function of the twist fields
$\phi^{+}(z_1,\bar{z}_1)$ (of \eq{ere}) on an (analytically extended)
strip ($\s_1 \in R, 0 \le \hbox{Re}(\s_2)  \le \b/2, 
\hbox{Im}(\s_2) =t >0$) with boundary conditions specified by the
state $| B \ran$. As before, we will map this geometry to the
(upper half) plane by using \eq{MapJ}, where the boundary condition
now applies to the boundary of the UHP. As discussed in 
\cite{Cardy:1984bb}, as long as the state $| B \ran$ represents
a conformally invariant condition (more on this in Section 
\ref{sec:info-loss}), the one-point function 
$\lan \phi^+(w_1, \bar w_1) \ran$ in the UHP is given by 
\be
\langle \phi^{+}(w_1,\bar{w}_1) \rangle|_{UHP}=
(w_1-\bar{w}_1)^{-h-\bar{h}}
\ee
which equals the two-point function of $\phi^+$ at $w_1$ 
with its image $\phi^-$ at $\bar{w_1}$ in the full plane.  
The original one-point function on the strip is now
obtained by \eq{MapJ}
\be \langle
\phi^{+}(z_1,\bar{z}_1)\rangle=\left(\frac{\beta_+}{\pi\,\epsilon}
\sinh\left(\frac{\pi\,z_1}{\beta_+}-\frac{\pi\,\bar{z}_1}{\beta_-}
\right)\right)^{-h}\left(\frac{\beta_-}{\pi\,\epsilon}
\sinh\left(\frac{\pi\,z_1}{\beta_+}-\frac{\pi\,\bar{z}_1}{\beta_-}
\right)\right)^{-\bar{h}}
\ee 
By putting the values \eq{z1-single}, we can compute the Renyi entropy.
Taking the $n\to 1$ limit, we obtain the EE  
\be
S_{EE}=\frac{c}{6}\log\left(\cosh\left(\frac{2\pi\,t}{\beta(1-\Omega^2)}
\right)\right)+\frac{c}{12}\log\frac{\beta^2(1-\Omega^2)}{\pi^2\epsilon^2}
\ee
For large $t \gg \b, \b\Om$, the EE evolves linearly with a
coefficient equal to half of the one for the thermofield double. For
$\Omega=0$ we recover the result of \cite{Hartman:2013qma}.

In the case of finite interval, let us suppose that the interval
stretches from $\s_1=-l/2$ and $\s_1= l/2$ at $\s_2=0$. At time
$t$ these end-points are translated to  $(z_1, \bar z_1)$ and $(z_2, \bar
z_2)$, where 
\begin{align}
z_1=-\frac{l}{2}+i\frac{\beta_+}{4}-t,\kern10pt \bar{z}_1
=-\frac{l}{2}-i\frac{\beta_-}{4}+t\nonumber, \qquad
z_2=\frac{l}{2}+i\frac{\beta_+}{4}-t,\kern10pt \bar{z}_2
=\frac{l}{2}-i\frac{\beta_-}{4}+t
\end{align}
The computation of the EE follows by using a slight modification of 
\cite{Calabrese:2005in,Hartman:2013qma}. The Renyi entropy
is given in terms of a two-point function on the above-mentioned strip
which can be obtained, from the UHP result
\be
\langle\phi_+(w_1)\phi_-(w_2)\rangle\sim\left(\frac{|w_1-\bar{w}_2|
|w_2-\bar{w}_1|}{|w_1-w_2||\bar{w}_1-\bar{w}_2||w_1-\bar{w}_1|
|w_2-\bar{w}_2|}\right)^{h+\bar{h}}
\ee
using the conformal transformation \eqref{MapJ}. The EE turns out to be
\be
S_{EE}=\frac{c}{6}\log\left(\frac{\beta_+\beta_-}{\pi^2\epsilon^2}
\frac{(\cosh\frac{2\pi\Omega\,l}{\beta(1-\Omega^2)}+\cosh
\frac{4\pi\,t}{\beta(1-\Omega^2)})\sinh\frac{\pi l}{\beta(1+\Omega)}
\sinh\frac{\pi l}{\beta(1-\Omega)}}{\cosh\frac{2\pi\,l}{\beta(1-\Omega^2)}+
\cosh\frac{4\pi\,t}{\beta(1-\Omega^2)}}\right)
\ee

\medskip
As in case of the thermo-field double, we again have, for large
$t/\beta$ and $l/\beta$, two cases (depending on the relative
magnitude of $t$ and $l/2$), that clearly illustrate the saturation of
the entanglement entropy (see Fig \ref{fig:Saturation})
\be
S_{EE}=\left \{       
\begin{array}{l l} 2t\, s_{\rm eqm}
   +S_{\rm div} & \quad t\le l/2\\
\\ 
l\, s_{\rm eqm}  
    +S_{\rm div}  & \quad t \ge l/2
  \end{array}          
                   \right.
\label{EE-J-single}
\ee \\ 
where $s_{\rm eqm}$ is the equilibrium entropy density given in
\eq{s-micro-J}. Note that the saturation  value of the entanglement
entropy for the single CFT is expectedly half of that in the case of
the thermofield double given by \eq{s-micro-J}. Also
note that the saturation value depends on the angular
momentum (see Fig \ref{fig:Saturation}).

The above equation \eq{EE-J-single} is again of the form of
\eq{thermal-chem}.  Thus, we have now proved this equation starting
from an arbitrary initial state \eq{b-state-eq}.

 \begin{figure}[!t]
       \begin{center}
     \includegraphics[scale=0.7]{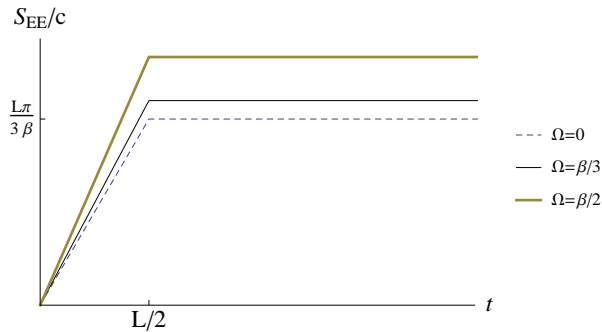} 
     \caption{\textit{Saturation of the Entanglement Entropy for different values of $\Omega$ }}
     \end{center}
     \label{fig:Saturation}
     \end{figure}

\subsection{\label{sec:info-loss}Information loss}

We wish to mention a rather remarkable feature of the EE described in
this subsection. By choosing the state $| B \ran$ in \eq{b-state-eq}
appropriately, we can make the initial state $| \psi \ran$ completely
arbitrary (contrast this with the state \eq{hartle-j} which is fixed
by the choice of $\b, \Om$); however, the entanglement entropy of an
interval in any such state is independent of the choice of the state
(this statement is even true for EE at any finite time).  The feature
of the calculation that makes this happen is the following. Recall
that the choice of $| \psi \ran$ corresponds to the choice of a
boundary condition for the two-dimensional CFT (in an appropriate
coordinate system, the state specifies a boundary condition on the
boundary of the upper half plane (UHP)). As has been shown in
\cite{Calabrese:2005in}, as long as the state $| \psi \ran$ is a
conformally invariant boundary state, the correlation function of
twist fields in the UHP, involved in computing the Reny entropy boils
down to correlators on the plane involving the original twist fields
and their images in the lower half plane. This result is universal and
is independent of the choice of the specific conformal boundary state,
of which there is an infinite tower (the so-called Ishibashi
states). Furthermore, as emphasized in \cite{Calabrese:2005in}, even
if our initial state is not one of the conformally invariant boundary
states, RG flow takes it to the nearby Ishibashi state; thus, for
sufficiently large length scales/time scales the result becomes
completely universal.  From the holographic viewpoint, the
universality is encapsulated by the fact that the bulk is given by a
BTZ black hole geometry. These features have already appeared in the
work of \cite{Hartman:2013qma}.  Such universalities with respect to
the initial state have also been remarked upon in
\cite{Balasubramanian:2011at,Liu:2013iza}.

\section{\label{sec:hol-J}EE with spin: holographic calculation}

As shown in \cite{Israel:1976ur, Maldacena:2001kr} (see also
\cite{Hubeny:2007xt,Cadoni:2009tk,Hartman:2013qma}) the above CFT
calculations find natural duals in BTZ geometries. For non-zero
angular momentum $J$, the holographic dual of the thermofield double
involves (a Euclidean continuation of) the eternal (2+1)-dimensional
BTZ black hole \cite{Banados:1992wn}, given by
\be 
ds^2= - \frac{(r^2-r^2_+)(r^2-r^2_-)}{r^2}dt^2+
\frac{r^2}{(r^2-r^2_+)(r^2-r^2_-)}dr^2+r^2
\left(d\phi - \frac{r_+r_-}{2\,r^2}dt\right)^2
\label{BTZJ}
\ee 
Here $\phi\sim \phi+2\pi$ for the BTZ black hole, and $\phi \in
{\mathbb{R}}$ for the BTZ black string. 
\footnote{\label{ftnt:black-string} The BTZ string can be obtained
  from the BTZ black hole by scaling $(r,r_\pm, t, \phi)$ with a
  parameter $\lambda$ such that, as $\lambda \to \infty$, $ \lambda r,
  \lambda r_\pm$, $t/\lambda, \phi/\lambda$ are held fixed. In this
  paper, we will mostly be concerned with the black string, since the
  dual CFT has non-compact space. For the black string, the angular
  momentum $J$ actually becomes the linear momentum; however, as
  declared in footnote \ref{ftnt:J}, we continue using the notation
  $J$ and the misnomer `angular momentum'.} The mass $M$, angular momentum
$J$, temperature $T$ and angular velocity $\Om$ are determined by
the inner ($r_-$) and outer ($r_+$) horizons, as follows:
\be
M=r^2_++r^2_-,\quad J=2\,r_+r_-, \quad 
T= 1/\b = \frac{2\pi r_+}{r_+^2 - r_-^2}, 
\quad \Om = \frac{r_-}{r_+}
\ee
The BTZ metric can be mapped into the Poincare patch of 
Euclidean $AdS_3$ ($ds^2= (dy^2 + dw_+ d w_-)/y^2$) via
\begin{align} 
w_\pm=\sqrt{\frac{r^2-r^2_+}{r^2-r^2_-}}\,e^{2\pi u_\pm/\b_\pm},\quad
y=\sqrt{\frac{r^2_+-r^2_-}{r^2-r^2_-}}\,e^{\pi\left(u_+/\b_+
+u_-/\b_-\right)}\label{map}
\end{align}
where $u_\pm=\phi\pm \,t$, and $\b_{\pm}=\b(1 \pm \Om)$ (see, e.g.,
\cite{KeskiVakkuri:1998nw}, Eq. (21)). The Euclidean continuation of
the above geometry \eq{BTZJ} is given by $t \to i t$, $u_\pm \to (z,
\bar z)$, $w_\pm \to (w, \bar w)$, $\Om \to i \Om_E$.  Note that in
the limit $r>>r_+$, the (Euclidean continuation of the) map \eq{map}
precisely reduces to the transformation \eqref{MapJ} in the CFT, as
it must for consistency with holography.

\medskip

We will now compute the holographic EE (hEE) for an interval $A$ by using
the Ryu-Takayanagi (RT) proposal \cite{Ryu:2006bv}, or more precisely
the generalization in \cite{Hubeny:2007xt} for computing covariant EE,
according to which the hEE is given by the length of the extremal
\footnote{The original RT
  prescription faces a subtlety for Lorentzian backgrounds. Namely, in
  general, geodesics that connect the boundaries do not lie on fixed
  time slices. In these cases EE is given by the area of the extremal
  surface given by the saddle point of the area action
  \cite{Hubeny:2007xt}.}  geodesic(s) that connects the boundary of
$A$. The precise formula reads
\be
S_{hEE}=\frac{\mL(\gamma)}{4G_{N}}
\label{ryu-t}
\ee 
where, in our case, $\mL(\gamma)$ is the length computed
with the metric \eqref{BTZJ} and $G_N$ is the Newton constant in 3
dimensions. Following along the lines of \cite{Hartman:2013qma}, it is
easy to verify that \eq{ryu-t} indeed reproduces the CFT results
\eq{s-micro-J} and \eq{EE-J-single}. We will skip the explicit
expressions since they are a straightforward generalization of
\cite{Hartman:2013qma} (we will make some more remarks on the geodesic
lengths below in Section \ref{summary}), and prefer to include an
alternative `derivation', which is more closely related to standard
AdS/CFT arguments.

\subsection{\label{sec:proof-ryu}A `proof' of Ryu-Takayanagi 
formula for 1+1 dimensional CFT}

 \begin{figure}[!t]
       \begin{center}
     \includegraphics[scale=0.7]{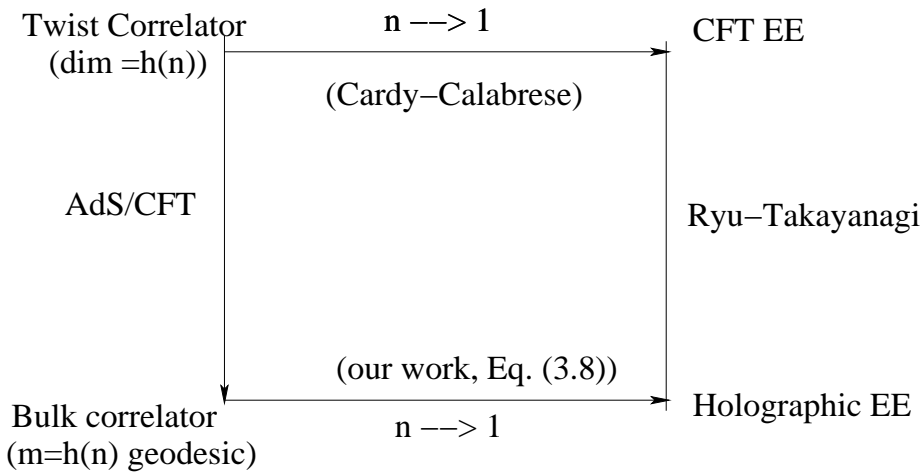} 
     \caption{A sketch of the `proof' of the Ryu-Takayanagi formula. The 
use of AdS/CFT map in the left vertical arrow is justified in the limit
of $n=1+ \epsilon$ (see text).}
     \end{center}
     \label{fig:comm-diag}
     \end{figure}

Recall that, in CFT, Entanglement Renyi Entropy (ERE) of a single
interval $[u,v]$ is computed by the two-point function of the twist
fields $\lan \Phi_+(u) \Phi_-(v) \ran$ with dimension\footnote{In the
  second equality we inserted the Brown-Henneaux relation
  $R_{AdS}/G_N=2c/3$. In our formulas $R_{AdS}=1$.}
\be
h=\frac{c}{24}\left(n-\frac{1}{n}\right)=\frac{1}{16G_N}
\left(n-\frac{1}{n}\right)
\label{DiM}
\ee
More precisely the ERE, from CFT, is given by
\be
S^{(n)}_{[u,v]}=\frac{1}{1-n}\ln \lan \phi_+(u) \phi_-(v) \ran
\label{ERT}
\ee
We will now show how in a double scaling limit  fixed,  the expression
\eq{ERT} reduces to the Ryu-Takayanagi expression \eq{ryu-t} through 
more or less standard AdS/CFT arguments.

Our strategy of computing \eq{ERT} would be to compute the CFT
two-point function holographically. We need to take the limit $c \to
\infty$ to ensure semiclassical gravity. The CFT two-point function
will be given in terms of a bulk propagator of a dual scalar field
whose mass $m$ (see the paragraph around 
\eq{double-scaling} for subtle assumptions
involved in the existence of such dual scalar fields), 
by the standard mass-dimension formula for large $c$, will be $m = h$. 
The bulk propagator between two points, on the other
hand, is given in terms of the geodesic length of a particle of mass
$m$ connecting the two points. Using these results,
the CFT two-point function in \eq{ERT} boils down to
\be
\lim_{c\to \infty} \lan \phi_+(u) \phi_-(v) \ran 
\xrightarrow{AdS_3/CFT_2} e^{-2h\,\mL(\gamma_{[u,v]}) }
\ee
where $\mL(\gamma_{[u,v]})$ is the length of the geodesic, in the
BTZ geometry, connecting the two boundary points $u$ and $v$.
Using this and \eq{DiM}, the holographic entanglement Renyi entropy 
\eq{ERT} reduces to 
\be
S^{(n)}_{[u,v]}=\frac{2h\,\mL(\gamma_{[u,v]})}{n-1}= 
\frac{\mL(\gamma_{[u,v]})}{8G_N} \left(1+1/n \right)
\label{proof-ERT}
\ee
Taking the  $n\to 1$ limit, we recover the formula 
prescribed by Ryu and Takayanagi \eqref{ryu-t}. 

The steps mentioned in the above `proof' are symbolically represented
in the `commutative diagram' in Figure \ref{fig:comm-diag}.

\paragraph{The importance of double scaling:}
In order to have a semiclassical gravity dual, we must take the limit
$c \to \infty$. Now, the conventional relation between two-point
functions of CFT primary fields and two-point functions of the
corresponding bulk duals assumes that dimensions of the CFT fields do
not scale with the central charge (this ensures the use of linear
response under deformation of the CFT by these fields). This
assumption appears to be, {\it a priori}, violated by the twist
operators with scaling dimensions \eq{DiM}. However, we
should recall that eventually we are interested in the entanglement
entropy which involves taking the limit $n \to 1$. What we
propose here is that we should take a judicious combination of
the $c\to \infty$ and $n\to 1$ limits; to be precise, let us
define the following double scaling limit
\begin{align}
c\to \infty,
n\to 1, c(n-1) = \hbox{fixed}
\label{double-scaling}
\end{align}
It is easy to see that in this limit the dimension $h$ of the twist
operator remains finite; hence computation of its two-point function
by the method described above should be justified.  In terms of the
commutative diagram Fig. \ref{fig:comm-diag}, the 
above remarks justify the use of AdS/CFT in
the left vertical arrow for $n= 1 + \epsilon$.
 
\medskip

A remark: it would be interesting to understand the connection of the
above argument with that of Lewkowycz and Maldacena in
\cite{Lewkowycz:2013nqa}. If one took the $c\to \infty$ limit without
the concurrent $n\to 1$, the bottom left of the commutative diagram in
Fig. \ref{fig:comm-diag} would be represented by the back-reacted
conical geometry sourced by the world-line of the very massive quantum
of the scalar field described above \cite{Deser:1983nh}; this is, at
least qualitatively, similar to the picture of
\cite{Lewkowycz:2013nqa}. However, in view of the above discussion, it
appears that in the double scaling limit described above, the bulk
partition function with the conical geometry is given in terms of a
propagator of the scalar particle in the undeformed geometry. We hope
to come back to this interesting issue in the future.

\subsection{\label{summary}Conclusion of this section}

In the light of our arguments above, the holographic computation and
its agreement with results of Section \ref{CFT-J} become very
transparent. In fact, to reproduce the CFT results now, we only need to
know the length of a massive geodesic (with mass equal to $h$) between
two points at the boundary of the spinning BTZ background. This length
was found in \cite{KeskiVakkuri:1998nw} (formula (34))\footnote{see
  also \cite{Louko:2000tp} for more discussion on the geodesic length
  and AdS/CFT correlators}. Using it in our algorithm precisely
reproduces the CFT two point function from bulk:
\be
\langle
\phi^{+}(z_1,\bar{z}_1)\phi^{-}(z_2,\bar{z}_2)\rangle=\left(\frac{\beta_+}
{\pi\,\epsilon}\sinh\frac{\pi\,l}{\beta_+}\right)^{-2h}
\left(\frac{\beta_-}{\pi\,\epsilon}\sinh\frac{\pi\,\bar{l}}{\beta_-}
\right)^{-2\bar{h}}\label{2ptTFD}
\ee
Since the CFT two-point function itself 
is reproduced, we get the same entanglement entropy as in the CFT.

For two finite intervals (which appear here for the thermofield double)
we use the same arguments since the four point correlator factorizes
into a product of two-point functions (see also
\cite{Hartman:2013mia}), which are then computed using the bulk
propagator, as above. Similarly, the holographic EE for the pure
B-state is just the half of the full space answer, as found earlier in
the $\Om=0$ case in \cite{Hartman:2013qma}).

We have thus holographically derived \eq{s-micro-J} and
\eq{EE-J-single} and hence holographically proved \eq{thermal-chem}
for the case with non-zero angular momentum.

\section{\label{CFT-JQ}EE for a charged state: CFT}

In this section, we will suppose that the CFT has a global $U(1)$
charge, and that the initial state has a non-zero value of this
charge. For simplicity, we will first consider the case of zero
angular momentum. In this case, the counterpart of \eq{hartle-j}
will be given by
\be
| \psi \ran = C \sum_i  \exp[- \beta  (E_i - \mu Q_i)/2]\, 
| i\ran \otimes  | i \ran
\label{hartle-q}
\ee
The $U(1)$ symmetry implies that the 
CFT has a $U(1)$ Kac-Moody algebra
\be
J(z)  J(0) =  k/(2z^2)  + {\rm regular~terms}
\label{level-k}
\ee
plus its antiholomorphic counterpart. The Kac-Moody
currents have the usual OPE with the stress tensor $T_{zz},
\bar T_{\bar z\bar z}$. 

\medskip
It is well-known, e.g. in the context of ${\cal N}=2$ superconformal
field theories, that the Kac-Moody and Virasoro algebras admit an
automorphism called ``spectral flow''. By choosing
a  flow parameter $\eta = \mu/2$, (using the conventions of
\cite{Kraus:2006nb}, Eq. (2.7))  we
find the following expression for the automorphism
\begin{align}
L_0 &\to L_0^{(\mu)} =  L_0 - \frac{\mu}{k} Q/2 +  \frac{\mu^2}{4k},\;
Q \to  Q^{(\mu)} = Q -  \mu/k
\label{spectral}
\end{align}
where $k$ is the level of the U(1) Kac-Moody algebra, defined in
\eq{level-k}. Although it is perhaps best studied in the
context of ${\cal N}=2$ superconformal theories, the phenomenon
of spectral flow is very generic; it exists  for
simple systems such as free massless charged fermions (see
Section \ref{sec:fermion}) for which half-integral spectral flows
connect the NS and R sectors; indeed in 
\cite{Gaberdiel:2012yb} arguments have been presented for its
appearance under rather general circumstances.

With this proviso, we will assume that the charged models
we have possess a spectral flow. It is easy, then, to
see that the CFT calculations in previous sections can be
simply generalized by using the unitary transformation
implementing the spectral flow. For example, consider 
the Renyi entropy
for the CFT on the plane which has the generic form
\be
S^{(n)}_{Renyi}=  Z_n/Z_1^n
\label{renyi}
\ee
where $Z_n$ is a partition function of an appropriate $n$-sheeted
surface and $Z_1$ is the partition function on the plane.
Now note that, by spectral flow (using $H= 2L_0$ for $J=0$),
\be
\Tr \exp[-\b (H - \mu Q + k\mu^{2}/2)] = \Tr \exp[-\b H],
\ee
which trivially leads to
\be
\Tr \exp[-\b (H - \mu Q)] = \Tr \exp[-\b (H - \mu^{2}/4k)]
\label{energy-shift}
\ee
Thus, the effect of adding the $\mu Q$ term is equivalent, in the
partition function and hence in \eq{renyi}, to the universal shift
\eq{energy-shift} to the Hamiltonian.  Using this line of reasoning,
it is easy to show that the time-dependent EE is given by applying
this shift to the energy $E$ in the expression for $s_{\rm eqm}$.  The
generalization to non-zero $J$ is straightforward, in that the same
shift again applies to the energy $E$. Using this shift to
\eq{s-micro-J}, and the relation between
$\mu$ and $Q$ as in Section \ref{sec:thermo} (this relation is
also discussed in \cite{Kraus:2006nb, Gaberdiel:2012yb}), we now get the
general result for the dynamical EE for non-zero $E, J$ and $Q$
\begin{align}
&S_{\rm ent}(t,l| \psi ) =
\left \{       
\begin{array}{l l}
\frac{t}{2}\,s_{\rm eqm}(E, J, Q) +S_{\rm div}, & \quad t\le l/2 \\
l\,s_{\rm eqm}(E, J, Q) +S_{\rm div}, & \quad t > l/2  \end{array}          
                   \right. 
\nonumber\\
& s_{\rm eqm}=  \sqrt{\frac{\pi c}{6}(E+J - \frac\pi{2k}Q^2)} + 
\sqrt{\frac{\pi c}{6}(E-J - \frac\pi{2k}Q^2)}
\label{s-micro-JQ}
\end{align} 
The dynamical EE for the pure state and single CFT with non-zero $Q$
follows similarly and it is given by half of the above result for the
thermofield double.

We have thus derived  the form of the dynamical
EE,  \eq{thermal-chem}, for arbitrary $E, J, Q$.  

\section{\label{sec:AdS-JQ}Holographic EE with charge: BTZ plus CS U(1)}

The bulk dual of the above CFT has been described in various places;
in particular, we will follow the account given in
\cite{Kraus:2006nb}.  The bulk dual consists of AdS gravity plus a
bulk $U(1)$ CS gauge field.  The metric is given, as in Sec
\ref{sec:hol-J}, by a spinning black hole \eq{BTZJ}; in addition,
there is a  bulk gauge field  solution given by the flat connection 
\be
A=  \frac{\mu}{k} (dz + d\bar z)
\label{a-sol}
\ee
Before proceeding, we now encounter an obvious puzzle.

\subsection{\label{sec:puzzle}A puzzle}

In the previous section (Section \ref{CFT-JQ}) we found that the
entropy density clearly depends on the charge $Q$.  The grand
canonical expression of such an entropy density is, therefore,
expected to be of the form $s(\b, \Om, \mu)$. In the bulk dual, as we
just mentioned, $\mu$ appears only in the gauge field solution
\eq{a-sol} and not in the metric which retains the $\mu=0$ form,
\eq{BTZJ}. The Ryu-Takayanagi prescription, therefore, will give the
time-dependent EE as in the uncharged case. In particular, we will
again obtain \eq{s-micro-J} and \eq{EE-J-single}. This clearly appears
to contradict \eq{s-micro-JQ} which we derived in the CFT.

Indeed, rather than for the EE, one could ask the same question about
the BH entropy. By the black hole area law, 
\be
s= s(\b)=\pi c/(3\b(1- \Om^2))
\label{area-law}
\ee
which is clearly independent of the chemical potential $\mu$
for the charge. Indeed, in terms of the microcanonical ensemble,
the above entropy density precisely agrees with \eq{s-micro-J}.
The puzzle is, how can we get the entropy density in
\eq{s-micro-JQ} from gravity?

\gap2

The resolution of this puzzle can be described in two equivalent ways,
one in the language of the microcanonical ensemble and the second in
the language of the (equivalent) grand canonical ensemble.

\begin{itemize}

\item \underbar{`Microcanonical' resolution}: 

Although the U(1) CS action in the bulk is topological and hence does
not couple to the metric, it has a boundary term of the form $A^2$
(see, e.g.  \cite{Kraus:2006nb}). This leads to an additional
contribution to $T_{zz}$ at the boundary, resulting in the following shift
\be
L_0 = L_{0,{\rm bulk}} +  Q^2/(4k), \; 
\bar L_0 = \bar L_{0,{\rm bulk}} +  Q^2/(4k),
\label{bulk-bdry}
\ee
This shift, in fact, is the bulk equivalent of \eq{energy-shift}.
See also Sec \ref{sec:Limits} for another application of this shift. 

\item \underbar{`Grand Canonical' resolution}: 
 
The microcanonical expression for the entropy density in
\eq{s-micro-JQ} can be converted into the grand canonical form using
the formulae in Section \ref{sec:thermo}.  Surprisingly, from that
expression, the $\mu$-dependence drops out, leaving the expression
\eq{area-law}!  The temperature and the two `chemical potentials'
$\Om, \mu$ are of course the same in the CFT and in the AdS dual;
hence we get agreement between the bulk and boundary expressions.

\end{itemize}

\underbar{Summary} We have thus proved \eq{thermal-chem}
holographically in the presence of both angular momentum and
charge.

\subsection{\label{sec:hs}Higher Spin}

It is natural to speculate how to extend the above calculations to the
case of further additional charges. A natural setting for this is to
consider higher spin black hole backgrounds whose CFT dual corresponds
to coset models \cite{Gaberdiel:2010pz} (this speculation 
was made earlier in \cite{Mandal:2013id}).  There exist limits of the
parameter space of this duality, which are described by free fermions
which describe a particularly simple form of an integrable CFT. A
similar integrable system of free fermions was recently discussed in
\cite{Mandal:2013id} where a version of \eq{ergodic} was found to be
true in the framework of the generalized Gibbs ensemble, and it was
speculated there that the equilibrium configuration of the
Gaberdiel-Gopakumar free fermions could be given by the higher spin
black holes. This makes it rather natural to conjecture that
\eq{thermal-GGE} should be true in this case, where the bulk dual
geometry should be that of a higher spin black hole. See the most recent progress in this direction \cite{deBoer:2013vca, Ammon:2013hba}.

\section{\label{sec:Limits}Universal limits}

By now, there is a lot of evidence that in the limit of a small entangling region\footnote{Or in the limit of $\beta\to \infty$}  $A$, EE obeys the analogue of the first law of thermodynamics \cite{Bhattacharya:2012mi,Nozaki:2013vta,Allahbakhshi:2013rda,Wong:2013gua}
\be
\Delta E_A=T_{ent}\,\Delta S_A\label{FL}
\ee
Here, the increase of energy in an interval $A=l$ is computed by integrating the holographic energy-momentum tensor $T_{tt}$ over the entangling interval, $\Delta S_A$ is a leading-$l$ difference between the EE computed in an excited state and that in the vacuum, and $T_{ent}$ is a universal constant that depends on the number of dimensions. More explicitly
\be
\Delta E_{A}=\int_{A=l}dx\,T_{tt}=\frac{MR\,l}{16\pi G_N}
\ee
where $R$ is the radius of asymptotically $AdS_3$ background with time component of the metric given by $f(z)^{-1}\sim 1+Mz^2$.

\medskip
For spinning BTZ solution \eqref{BTZJ}, the increase EE of a single interval of length $l$ \cite{Hubeny:2007xt} is given by 
\bea
\Delta S_l=\frac{c}{6}\log\left(\frac{\beta_+\beta_-}{\pi^2\epsilon^2}\sinh\frac{\pi l}{\beta_+}\,\sinh\frac{\pi l}{\beta_-}\right)-\frac{c}{3}\log\frac{l}{\epsilon}\sim \frac{c\pi^2(1+\Omega^2)T^2\,l^2}{18(1-\Omega^2)^2}
\eea
Using the relation between the mass and the temperature 
\be
M=(2\pi T)^{2}\frac{1+\Omega^2}{(1-\Omega^2)^2}
\ee  
the first law relation for EE becomes
\be
\Delta E_l=\frac{3}{\pi l}\Delta S_l
\ee
in agreement with \cite{Bhattacharya:2012mi}.\\
On the other hand, in the limit of large $l$ or large temperature $T$, 
the EE reaches the extensive form given by the thermal entropy of the system 
(see the review \cite{Calabrese:2009qy})
\be
S_{l}\approx  l\,s_{\rm eqm}(\beta) \label{TH}
\ee
For spinning BTZ, $s_{\rm eqm}$ can be read off
from \eq{s-micro-J}. Thus, 
\be
S_l  \approx \frac{c\pi\,l}{3\beta(1-\Omega^2)}
\ee
The result depends only on the central charge of the CFT and
on $\b$ and $\Om$, and is, therefore, a universal limiting value.

\medskip
Let us now look how the universal limits incorporate the presence of the U(1) CS fields. As explained in the previous section, U(1) gauge fields give an additional boundary contribution to the energy-momentum tensor
\be
T_{tt}=T^{\rm grav}_{tt}+T^{\rm gauge}_{tt}=\frac{M R}{16\pi G_N}+\frac{\mu^2}{2\pi}=\frac{\pi c}{6\beta^2}+\frac{\mu^2}{4\pi k}\equiv E_{\rm bulk}+\frac{\mu^2}{4\pi k}
\ee
This is consistent with (5.4) noting $T_{tt}=L_0/\pi$. This way, using the spectral flow argument, we have, as in \eq{bulk-bdry} 
\be
E_{\rm bulk}=E_{\rm bdry}-\frac{\mu^2}{4\pi k}
\ee
and the first law-like relation remains the same.

\medskip
The value of the thermal entropy at which EE saturates can be
expressed in terms of microcanonical energy density $E$ and potential
$\mu$. However, in the grand canonical ensemble it is only a function
of $\beta$ that matches the holographic prescription that is``blind"
to the gauge fields, as we noted in the previous section.

\section{\label{sec:erg}Relation with quantum ergodicity}

The idea of ergodicity is that given sufficient time, the ``time
average of various properties of a system'', evolving from some
initial state $\mathscr{S}_0$ can be equated to an ``ensemble average
of those properties'', where the ensemble is constructed out of all
possible states $\mathscr{S}$ of the system which have the same
conserved charges as $\mathscr{S}_0$. In the classical version, the
initial state is a point in the phase space; ergodicity says that
under dynamical evolution the point moves ``democratically'' in the
submanifold $M$ of the phase space, allowed by conservation laws, so
that the time average of a phase space function $f(q,p)$ can be
equated to an average of $f$ taken over $M$ with uniform weight. 
In quantum mechanics, under the usual conservation law
of energy, the statement boils down to 
(see, e.g.,  the review in \cite{Polkovnikov:2010yn})
\bea
(1/T)\int_0^T dt\  \Tr\ \left(\rho_{\rm pure}(t)\ O
\right)
\xrightarrow{T\to \infty}
\Tr \left(\rho_{\rm micro}\ O\right),  \quad   \rho_{\rm pure}
= | \psi \ran \lan \psi |, \; \nonumber\\
\rho_{\rm micro} = \frac1N \sum_{i=1}^N |i\ran \lan i|
\approx \rho_{\rm thermal} = 
\frac1Z \exp[-\beta H]
\label{ergodic-micro}
\eea
Sometimes an alternative statement (Eq. \eq{ergodic})  is made 
\cite{Polkovnikov:2010yn}
\be
\Tr \left( \rho_{\rm pure}\ O \right)   \xrightarrow{t\to \infty}
\Tr \left( \rho_{\rm thermal}\ O \right)
\label{ergodic-2}
\ee
with respect to a certain class of ``macroscopic'' observables, for
which the time averaging in \eq{ergodic-micro} is not
necessary. Let us now consider a partition $A \cup B$ of space,
say for quantum field theory, or for spins on a lattice.
Consider a basis of states $| i_A \ran |n_B \ran$ where
the states $| i_A \ran$ are supported entirely on $A$, and the
states $| n_B \ran$ are supported entirely on B (it could consist
of spins in $A$  and $B$. Let us consider the projection
operator $P_B = \sum_n |n_B\ran \lan n_B |$ onto the states
of $B$. To make connection with the discussion above, we 
choose $O= P_B\ | i_A \ran \lan j_A |$ in \eq{ergodic}, assuming
that this is an appropriate ``macroscopic'' operator. 
Eq. \eq{ergodic} then gives us  the following limiting value of
the matrix element of the reduced density matrix $\rho_A
= \Tr ( P_B \ \rho_{\rm pure})$
\begin{align}
\lan i_A | \rho_A | j_A \ran \xrightarrow{ t\to \infty}
\lan i_A |\rho_{A, \b}| j_A \ran, \qquad
\rho_{A, \b} \equiv \Tr \left(P_B \ \exp[-\b H]/Z \right)
\label{def-rhoab}
\end{align}
The asymptotic value of the time-dependent EE, then would be
\be
S_{\rm EE}  \xrightarrow{ t\to \infty}
S[\rho_{A, \b}] \equiv  - \Tr \left( 
\rho_{A, \b} \ln \rho_{A, \b} \right)
\label{s-rhoa-beta}
\ee
The latter entropy measures the von Neumann entropy
of the reduced density matrix {\it in an overall mixed state}.
Now we expect that for a large enough $l= l_A$ 
\be
S[\rho_{A, \b}]/l = s(\b)
\label{reasonable}
\ee
where $s(\b)$ is the thermal entropy density at 
an inverse temperature $\b$. In Section \ref{sec:proof}
we present a proof of this statement using a discrete
system and assuming the equivalence between microcanonical
and grand canonical ensembles. Additionally, 
in Section \ref{sec:massive}
we  explicitly verify \eq{reasonable} in the case
of a massive, charged scalar field. 

We have therefore proved 
\be
S_{\rm EE}  \xrightarrow{ t\to \infty}
l \, s
\label{thermal-ergodic}
\ee
This is the same as \eq{thermal}, where a time scale
of saturation is set by $l$. 

The proof of saturation outlined
above holds in principle for any field theory
and in any number of dimensions. Hence, we expect the
behaviour \eq{thermal-ergodic} to be valid quite
generally. 

For integrable systems and the generalized Gibbs
ensemble, the story of quantum ergodicity is less developed,
although we still expect an equation of the form \eq{ergodic-2}
to hold for a suitable class of ``macroscopic'' observables
(see \cite{Polkovnikov:2010yn}). \footnote{These
observables typically display some
non-locality; however, see \cite{Caux:2013} for the behaviour
of local observables.} 

\section{\label{sec:massive}Non-CFT: EE for charged, massive scalar field}

In this section, we consider an {\it a priori} calculation of EE for a
charged, massive scalar field. The motivation for this calculation
is to have an additional evidence for \eq{reasonable}.
Our CFT calculations for the saturation value of the time-dependent
EE already provide indirect evidence for this formula. However,
in this section we consider a non-conformal system and perform
a direct computation of the EE using the methods of \cite{Herzog:2012bw}.

\begin{figure}[t]
\centerline{\includegraphics[scale=0.5]{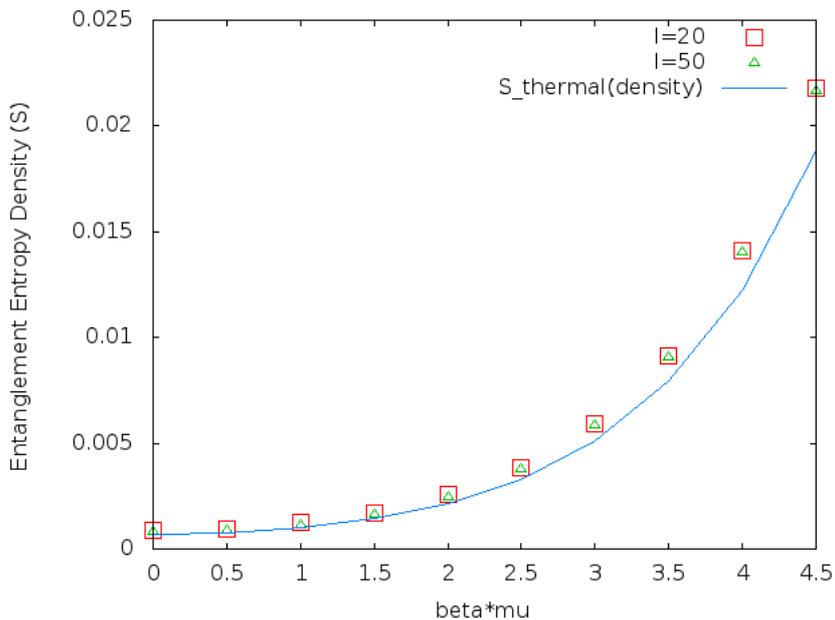}}
\caption{Plot of  $S[\rho_{A,\beta}]/l$  (the
LHS of \eq{reasonable} vs $\beta\mu$, for $\beta m=10$.
The plot marked with squares has $l/\beta=10$; the plot
marked with triangles has $l/\beta=25$. The solid line corresponds
to the RHS of \eq{reasonable}, viz. the thermal entropy
density, which is obtained using standard formulae, and Mathematica.
It is clear that \eq{reasonable} holds to a good accuracy.
The agreement is better for small $\beta\mu$
than for large $\beta\mu$; however, this could be due to
some numerical instability.} 
\label{ent-betamu-fig}
\end{figure}

The charged scalar field is described by a Hamiltonian
\begin{equation}
  H = \int dx \left( \pi^\dagger\pi + (\nabla\phi)^\dagger(\nabla\phi) + m^2
  \phi^\dagger\phi \right)
\end{equation}
and a conserved U(1) charge  
\be
Q= i \int dx (\pi^\dagger \phi -  \phi^\dagger \pi)
\ee
We will suppose that the full system (with spatial partition $A$
and its complement $B$) is in a grand canonical ensemble
\begin{equation}
\rho_{\rm total} =\frac{\exp [-\beta(H-\mu Q)]}{\Tr[ \exp [-\beta(H-\mu Q)]]}  
\end{equation}
We are interested in computing the reduced density matrix $\rho_A =
\Tr_B \rho_{\rm total}$, and the EE $S_A = - \Tr \rho_A \ln \rho_A$.
We will proceed using a generalization of 
the formalism described in \cite{Herzog:2012bw},
and describe only the essentially new features. Note that formalism in
\cite{Herzog:2012bw} permits a straightforward generalization
from the case of a single scalar $\phi$ to multiple flavours
$\phi_a, a=1,2,..., N_f$, with the $C$-matrix generalized to
\begin{equation}
  (C^2)_{ij}^{ab} = \sum_{k=1}^{n} <\phi_i^a\phi_k^c><\pi_k^c\pi_j^b>
\label{c-matrix}
\end{equation}
The EE is given by (using the notation from \eq{s-rhoa-beta})
\begin{equation}
  S[\rho_{A,\beta}]= \Tr[(C + 1/2)\log(C+1/2) - (C-1/2)\log(C-1/2)]
\label{herzog-ee}
\end{equation}
where the trace is now over both the $\{i,j\}$ and $\{a,b\}$ indices.
For the free complex scalar at hand, $N_f=2$, and $C_{ij}^{ab}
= C_{ij} \delta^{ab}$. We compute $C^2$ using \eq{c-matrix}
and the following ingredients:
\begin{align}
&<\phi_j\phi_k^\dagger> =\frac{1}{2N} \sum_{a=0}^{N-1} \frac{1}{2\epsilon\omega_i} [\coth(\frac{\omega_i+\mu}{2T})+\coth(\frac{\omega_i-\mu}{2T})] \cos [\frac{2\pi}{N} a(j-k)]  
\nonumber\\
& <\pi_j\pi_k^\dagger> = \frac{1}{2N} \sum_{a=0}^{N-1} \frac{\epsilon\omega_i}{2} [\coth(\frac{\omega_i+\mu}{2T})+\coth(\frac{\omega_i-\mu}{2T})] \cos [\frac{2\pi}{N} a(j-k)]  
\nonumber\\
&<\pi_j\pi_k^\dagger>= <\pi_j^\dagger\pi_k>,\quad
    <\phi_j\phi_k^\dagger>= <\phi_j^\dagger\phi_k>
\end{align}
The computation of \eq{herzog-ee} is performed numerically. We
reproduce a representative plot in Fig. \ref{ent-betamu-fig}.

\medskip
\subsection*{Acknowledgments}

It is a pleasure to thank Pallab Basu, Justin David, Abhishek Dhar,
Matthew Headrick, Sachin Jain, Nilay Kundu, Shiraz Minwalla, Takeshi
Morita, Johannes Oberreuter, Aninda Sinha, Nilanjan Sircar and Sandip
Trivedi for very useful discussions and comments. We are grateful to
Dileep Jatkar, Mukund Rangamani, Aninda Sinha and 
Spenta Wadia for important feedbacks on earlier versions of 
the preprint. GM would like to
thank the organizers and participants of (a) the ICTS meeting
``US-India advanced studies institute on thermalization: from glasses
to black holes'' held at IISc, Bangalore (June 10-21, 2013), and (b)
the ``Seventh Crete Regional Meeting on String Theory'' held at
Kolyambari, Crete, Greece (June 16-22, 2013), for hospitality and
stimulating discussions during the finishing stage of this work. GM
would also like to thank the organizers of the latter meeting for an
opportunity to present this work. The work of PC is based upon
research supported by the South African Research Chairs Initiative of
the Department of Science and Technology and National Research
Foundation. PC would also like to thank Tata Institute for Fundamental
Research in Mumbai for hospitality and support during this project.

\appendix
\section{\label{sec:thermo}Microcanonical vs grand canonical quantities}

Consider a grand canonical ensemble given by the 
following density matrix and partition function
\be
\rho = (1/Z)\exp[-\b(H + \Om P - \mu Q)],\;
Z= \Tr \exp[-\b(H + \Om P - \mu Q)]
\label{gce}
\ee
The partition function can be written as
\be
Z \equiv
\exp[-\b G]= \sum_{E,J,Q} \exp[S(E,Q,J)-\b(E + \Om P - \mu Q)]
\ee
If the summand in the last function has a single sharp maximum
around a unique set of values $E, P, Q$, the distribution
essentially becomes equivalent to that of a microcanonical ensemble, where
we have
\be
\del S/\del E= \b,\qquad \del S/\del J=- \b \Om, \qquad 
\del S/\del Q= - \b \mu
\label{micro-gce}
\ee
This gives us the grand canonical parameters in terms of the microcanonical
ones. The converse relations are also easy to derive:
\be
\del (\b G)/\del \b= E- \Om J - \mu Q,~  -1/\b ~\del (\b G)/\del \Om= J,~
-1/\b ~ \del (\b G)/\del \mu= Q
\label{gce-micro}   
\ee
This gives us the following relation
\be
S= \b^2 \del G/\del \b
\label{gce-s}
\ee
Using the above relations, we can prove that if the microcanonical
entropy density is given by
\be
S= \sqrt{a E +  b J - d Q^2}  +  \sqrt{a E - b J - d Q^2}
\label{s-micro}
\ee
the grand canonical expression for the entropy is
\be
s= \frac{2 a}{\b (1-  a^2\Om^2/b^2)}
\label{s-gce}
\ee
Surprisingly, $S(\b, \Om, \mu)=  S(\b, \Om)$, which is
independent of $\b$. 

\underbar{Notation}: In the body of the paper, we have used the
notations $E, J, Q$ as the energy {\it density}, $J$ as the angular
momentum {\it density} and $Q$ as the charge {\it density} whereas $s$
denotes the entropy density. Equations \eq{s-micro} and \eq{s-gce}
hold for the densities with trivial modifications.

\subsection{\label{sec:fermion}Free massless charged fermion in 1+1}

We consider free massless charged fermions in 1+1 dimension, at a
temperature $1/\beta$ and chemical potential $\mu$.
Explicit calculation gives the following Gibbs free energy 
\be
g(\b, \mu)=  - \left( \frac\pi{6\b^2}+ \frac{\mu^2}{2\pi} \right)
\label{g-fermions}
\ee
The energy density $e$ and the charge density $q$ are given by
\be
e=   \frac\pi{6\b^2} + \frac{\mu^2}{2\pi},  ~  
q = \frac\mu\pi  
\ee
For this system the grand canonical entropy density is given by
\be
s(\b, \mu)= s(\b)= \frac\pi{3 \b}
\label{s-gce-fermi}
\ee
and the microcanonical entropy density is given by
\be
s(e,q)=  \sqrt{\frac{2\pi}{3}(e -  \frac\pi{2} q^2)}
\label{s-micro-fermi}
\ee

\section{\label{sec:proof}Proof of \eq{reasonable}}

Consider a basis of the Hilbert space $\{ |i, A \ran, 
| k, B \ran \}, i=1,..., N_A; k=1,...N_B$, where $N_A,
N_B$ denote the number of independent states $|i, A\ran$
belonging to the partition A, and similarly for $N_B$. 
A microcanonical density matrix $\rho_{\rm mc}$
is given by
\be
\rho_{\rm mc}= \frac1{N_A N_B}\sum_{i,A; k,B} |i, A \ran  
| k, B \ran \lan i, A| \lan k, B \ran
\label{rho-mc}
\ee
By tracing over the B states, we get
\[
\rho_{A, mc} \equiv \Tr_B \rho_{\rm mc} = \frac1{N_A}\sum_{i,A} |i, A \ran  
\lan i, A| 
\]
The von Neumann entropy of this density matrix is given by
\[
S_A = \ln N_A
\]
Now imagine that our system is a lattice of $n$-level `spins'
($n=2$ is Ising), and that there are $l_A$ spins in the
partition A; then 
\[
S_A =  l_A \ln n
\]
Now we can easily show that the von Neumann entropy of 
\eq{rho-mc} is 
\[
S=  (l_A + l_B) \ln n
\]
Hence the entropy density is  
\[
s = \ln n
\]
Now by denoting $l_A$ as $l$, and assuming the 
equivalence between the microcanonical ensemble in 
\eq{rho-mc}  and the canonical ensemble as in \eq{def-rhoab}
we obtain \eq{reasonable}.


\end{document}